\documentclass[english]{article}
\usepackage[T1]{fontenc}
\usepackage[latin9]{inputenc}
\usepackage{geometry}
\usepackage{array}
\usepackage{multirow}
\usepackage{setspace}

\makeatletter

\providecommand{\tabularnewline}{\\}

\usepackage[margin=10pt,font=small,labelfont=bf,labelsep=endash]{caption}
\usepackage{microtype}
\usepackage{lineno}
\usepackage{mathabx}
\date{} 

\makeatother

\usepackage{babel}
\begin{document}

\title{Funding the search for extraterrestrial intelligence with a lottery
bond}

\begin{singlespace}

\author{\noindent {\normalsize{}Jacob Haqq-Misra}\\
{\normalsize{}}\\
{\normalsize{}Blue Marble Space Institute of Science}\\
{\normalsize{}1001 4th Ave, Suite 3201, Seattle, Washington 98154,
USA}\\
{\normalsize{}jacob@bmsis.org}\\
{\normalsize{}}\\
{\normalsize{}Published in }\emph{\normalsize{}Space Policy}{\normalsize{},
doi:10.1016/j.spacepol.2017.03.003}}
\end{singlespace}
\maketitle
\begin{abstract}
I propose the establishment of a SETI Lottery Bond to provide a continued
source of funding for the search for extraterrestrial intelligence
(SETI). The SETI Lottery Bond is a fixed rate perpetual bond with
a lottery at maturity, where maturity occurs only upon discovery and
confirmation of extraterrestrial intelligent life. Investors in the
SETI Lottery Bond purchase shares that yield a fixed rate of interest
that continues indefinitely until SETI succeeds---at which point a
random subset of shares will be awarded a prize from a lottery pool.
SETI Lottery Bond shares also are transferable, so that investors
can benefact their shares to kin or trade them in secondary markets.
The total capital raised this way will provide a fund to be managed
by a financial institution, with annual payments from this fund to
support SETI research, pay investor interest, and contribute to the
lottery fund. Such a plan could generate several to tens of millions
of dollars for SETI research each year, which would help to revitalize
and expand facilities such as the Allen Telescope Array. The SETI
Lottery Bond is a savings product that only can be offered by a financial
institution with authorization to engage in banking and gaming activities.
I therefore suggest that one or more banks offer a lottery-linked
savings product in support of SETI research, with the added benefit
of promoting personal savings and intergenerational wealth building
among individuals.\textbf{}\\
\textbf{}\\
\textbf{Keywords}: SETI; science funding; savings bonds; lotteries;
long-term finance
\end{abstract}

\section{Introduction}

Finding sources of funding for the Search for Extraterrestrial Intelligence
(SETI) has proven to be almost elusive as ETI themselves, and the
lack of adequate funding is perhaps the greatest obstacle to the SETI
program's success. Evidence of ETI in the galaxy could take the form
of radio signals from other star systems {[}1{]}, interstellar spacecraft
sent to our Solar System {[}2, 3{]}, or infrared excesses generated
by space faring super-civilizations {[}4{]}, but the limited number
of searches for any such phenomenon so far has produced null results
{[}5{]}. From Senator William Proxmire's issuance of the ``golden
fleece award'' for NASA's sponsorship of SETI research in 1978, to
the financial shortfall that put the Allen Telescope Array into hibernation
in 2011, the SETI research community has suffered from lack of funding
since its inception---and for the obvious reasons that SETI represents
a game in which both risk and rewards are unknown. 

Most funding for SETI research takes the form of grants and donations,
often scant in supply and limited in duration. For SETI to succeed,
astronomers must scan the sky in search of directed or accidental
transmissions that indicate extraterrestrial intelligent life. To
avoid missing any such signals, the SETI program ideally requires
a total sky survey at all radio frequencies and at all times of day---a
feat that could take anywhere from a few years to a few thousand or
more. As a step toward this goal, Microsoft's co-founder Paul Allen
has donated over \$30 million toward the construction of the Allen
Telescope Array (ATA), a radio interferometer designed especially
for SETI, but this flagship facility is only partially operational
and still incomplete due to the lack of donors interested in funding
its continual operation. Likewise, few federal and private granting
agencies show interest in funding SETI research or supporting the
ATA, while any grants that do succeed only last a few months to a
few years. The SETI community has been relatively unsuccessful at
attracting investment capital, due in part to the SETI program's inability
to promise a return on investment to prospective investors. The discovery
of ETI could be one of the most philosophically provocative and socially
stimulating discoveries of human history {[}6, 7, 8{]}, yet this discovery
itself will do little to refill the pockets of investors who provide
long-term venture capital. As charitable research funding becomes
increasingly limited, novel sources of financing are needed to ensure
a sustainable future for the SETI program.

Here I propose to attract long-term investors in SETI research by
creating a ``SETI Lottery Bond'' (SLB) savings product that reaches
maturity only upon the first discovery and confirmation of ETI. Savings
products with a randomized return historically have been issued to
finance projects with low investor zeal, and contemporary examples
of such products include the U.K. Premium Bond program {[}9, 10, 11{]}
and lottery-linked/prize-linked deposit accounts used in a variety
of nations {[}12, 13{]}. The SLB is a product issued by a financial
institution, and the funds collected from the SLB are managed to provide
a fixed rate of return to the investor, a steady stream of funding
to support SETI research, and a contribution to a long-term lottery
prize. Investors in the SLB purchase bonds at a fixed price in exchange
for a guaranteed rate of interest for the lifetime of the bond and
are allowed to continue purchasing bonds until the first discovery
(and subsequent confirmation) of ETI. After this initial discovery
of ETI, SLB shares will be redeemed, thereby repaying the initial
investment and closing any secondary markets for trading SLB shares.
In addition, a lottery prize will be awarded randomly to a subset
of SLB shares upon discovery of ETI as an incentive to encourage long-term
investment and promote public enthusiasm for SETI. The SLB in effect
acts as a perpetual bond (such as a Consol) that yields interest as
income but is unlikely to be redeemed at any time in the near future.
Such a savings product may appeal to a wide range of investors and
would provide a financial backbone to sustain the long-term efforts
required for the SETI program's first success.

This proposal may seem provocative to some, but bonds and lotteries
have been successful means of financing otherwise unattractive investments,
and I suggest that SETI research could benefit from such an approach.
I begin in Section \ref{sec:Bonds-and-Lotteries} by defining the
concepts of bonds, perpetual bonds, lotteries, and lottery bonds that
are required for describing the SLB. In Section \ref{sec:SETI-Lottery-Bond}
I present a proposal for the SLB as a fixed rate perpetual bond with
a lottery at maturity and show that such a fund could generate several
to tens of millions of dollars for SETI research each year. I then
discuss some of the ethical considerations involved in funding research
with such a product in Section \ref{sec:Ethical-Considerations},
and conclude with prospects for future implementation.

\section{Bonds and Lotteries \label{sec:Bonds-and-Lotteries}}

The SETI Lottery Bond incorporates elements of perpetual bonds, a
savings product that never reaches maturation, and lotteries, a form
of gaming that involves the random selection of prizes to a subset
population. This section defines basic concepts of bonds, perpetual
bonds, lotteries, and lottery bonds in order to establish a framework
for discussing the SLB.

A \emph{bond} is a debt security that obligates the issuer to pay
a certain amount of interest to the holder at regular intervals with
a promise to repay the principal (the net purchase price of the bond)
at some point in the future. The time between a bond's issuance and
its redemption is known as the \emph{bond duration}, at which point
the bond is said to have reached \emph{maturity}. Bonds are often
offered by national governments to finance otherwise unattractive
investments (such as large construction projects or military operations),
usually with a bond duration on the order of years to decades. Many
bonds can also be bought and sold as commodities in a secondary market
that facillitates trading, although some bonds prohibit transfer of
ownership. Bonds are attractive savings products to investors because
they offer a relatively predictable return on investment over a known
duration of time.

While most bonds are issued with a fixed and known duration, \emph{perpetual
bonds} offer regular payments of interest that continue forever but
with no promise of ever repaying the principal investment. A true
perpetual bond therefore acts as an annuity that guarantees the purchaser
a fixed stream of income for life or longer. A \emph{Consol} is a
famous example of a perpetual bond that was first issued by the British
government in 1751 as an attempt to consolidate its outstanding debt.
Originally offering a 3.5\% return to investors, this interest rate
has declined through several subsequent conversions, and Consols today
in the U.K. offer a lower rate of 2.5\%. The duration of a Consol
bond is unknown and taken to be infinite by many investors. Technically,
Consols can be redeemed at any point in the future by an Act of Parliament,
but such an Act seems improbable in the current economic climate.
Barring such an unlikely Act, Consols represent one of the few examples
of perpetual bonds and provide a financial product similar to annuities.
Because a Consol is unlikely to ever be redeemed, an investor never
will be repaid their principal investment, but they will continue
to receive interest payments on their investment for life---and often
these bonds can be benefacted to kin or charity. Perpetual bonds are
attractive savings products for long-term investors seeking a regular
stream of income from their principal.

A \emph{lottery} is a form of gaming that provides a way to allocate
prizes or resources to a large population, usually by random selection
of a subset population. Lotteries are usually self-funded, in that
the value of the prize pool is determined by the number of lottery
tickets sold. A typical lottery will sell tickets at a fixed price
for a known duration of time, after which ticket sales are suspended
and winners randomly chosen. Lotteries also have been suggested as
a fair way of distributing scarce resources, such as selecting patients
to receive an organ transplant {[}14, 15{]}. Many state governments
in the U.S. operate lotteries to support social programs such as public
education, economic development, ecological preservation, and addiction
treatment, although some evidence suggests that individuals who purchase
state lottery tickets are less likely to receive direct benefits from
lottery funds {[}16, 17{]}. Lotteries tend to attract a disproportionate
number of low- to middle-income households {[}16, 17{]} and appeal
to optimistic and adventuresome investors who hope to achieve a large
return on their investment. 

A \emph{lottery bond} is a savings product that combines elements
of bonds and lotteries. Lottery bonds offer repayment of the principal
investment like a conventional bond, but they also include a lottery
prize that is awarded to a subset of shares periodically or when the
bond reaches maturity. Investors in a lottery bond opt for a lower
rate of interest (compared to a conventional bond) in exchange for
a chance at winning a much larger prize. However, unlike a conventional
lottery, investors in a lottery bond do not risk loss of their principal
investment; lottery bonds instead represent a savings product that
incorporates a randomized rate of return. Lottery bonds have substantial
historical precedent and have been issued in the past and present
by many European nations {[}18, 19{]}. The U.K. Premium Bond program
is a contemporary example of a government-issued lottery bond that
provides monthly drawings for cash prizes, with an investor's chances
of winning proportional to the number of bonds owned {[}9, 10, 11{]}.
Other examples of similar financial products include lottery-linked
deposit accounts {[}12{]} and prize-linked savings accounts {[}13{]},
both of which provide depositors a low rate of interest on their account
balance along with a periodic chance at winning a much larger prize.
Prize-linked and lottery-linked savings products are used in a variety
of developing and industrialized nations and have shown some success
at encouraging fiscal responsibility among individuals otherwise likely
to gamble {[}12, 13, 20{]}. Lottery bonds are attractive savings products
to adventuresome investors who seek a large rate of return but also
want to retain their principal investment.

\section{SETI Lottery Bond\label{sec:SETI-Lottery-Bond}}

The purpose of the SETI Lottery Bond is to generate a source of revenue
that can sustain the SETI program until the first discovery of extraterrestrial
intelligent life. By combining elements of perpetual bonds and lotteries,
the SETI Lottery Bond is a unique savings product that also supports
scientific research.

Most conventional sources of funding stipulate a finite duration of
time for a project to reach completeness; however, it is nearly impossible
to calculate the probability of success for SETI, which creates great
difficulty in predicting the expected amount of time required for
SETI to succeed. A conventional bond therefore is an inadequate device
for funding SETI because bonds require the lender to repay the principal
at a specified future time---even a thousand year duration for a bond
might be insufficient time for SETI to succeed. Perpetual bonds provide
a better mechanism for funding SETI because the duration of a perpetual
bond essentially is unlimited: a perpetual bond linked to the outcome
of SETI would pay investors a fixed rate of interest and only return
the principal investment upon the first success of SETI. In principle,
such a perpetual bond might be sufficient to attract long-term investments
in SETI; but in practice, such a savings product would have few advantages
over other similar products already offered by savings institutions,
such as annuities and existing perpetual bonds. Thus, in order to
incentivise a perpetual bond while simultaneously promoting awareness
of SETI research, I propose the creation of a ``SETI Lottery Bond''
that acts as a perpetual bond but with a lottery prize awarded to
a random subset of investors upon SETI's first discovery.

The SETI Lottery Bond is defined as a fixed rate bond with a lottery
at maturation, where maturation occurs upon the discovery of extraterrestrial
intelligent life. Let $r_{c}$ be the fixed rate of return on investment
offered by the SLB (also known as the \emph{coupon rate}), let $P$
be the price at which each share in the SLB is sold, and let $N$
be the total number of SLB shares sold. The total capital raised is
therefore $NP$ and the amount owed to investors each year is $r_{c}NP$.
Additionally, the SLB pays a yearly contribution $r_{l}$ toward the
lottery fund along with a yearly contribution $r_{s}$ toward SETI
research, which amounts to $(r_{l}+r_{s})NP$ each year. Assuming
a constant rate of inflation $r_{i}$ and annual compounding, and
letting $t$ represent time, the future value $V_{future}$ of each
share of the SLB can be expressed as

\begin{equation}
V_{future}=P\left(\frac{1+r_{c}}{1+r_{i}}\right)^{t}.\label{eq:FV}
\end{equation}
If the rate of inflation is greater than the bond coupon rate ($r_{i}>r_{c})$,
which is likely the case for a lottery-linked product, then the future
value $V_{future}$ should decrease with time. However, the expected
value $V_{expected}$ of each share takes into account the possibility
of winning a lottery prize when the bond matures. The expected value
can be expressed as

\begin{equation}
V_{expected}=V_{future}+r_{l}Pt,\label{eq:EV}
\end{equation}
which increases with time as the lottery prize fund grows. An example
of SLB prices is shown in Table \ref{tab:Prices}
\begin{table}
\begin{centering}
\begin{tabular}{cccc}
\hline 
\multirow{1}{*}{\textbf{Time (yr)}} & \multirow{1}{*}{\textbf{Prize Fund}} & \textbf{Future Value} & \textbf{Expected Value}\tabularnewline
\hline 
1 & \$0.75M & \$99.03 & \$99.78\tabularnewline
3 & \$2.25M & \$97.12 & \$99.37\tabularnewline
6 & \$4.50M & \$94.31 & \$98.81\tabularnewline
10 & \$7.50M & \$90.70 & \$98.20\tabularnewline
30 & \$22.5M & \$74.63 & \$97.13\tabularnewline
60 & \$45.0M & \$55.69 & \$100.69\tabularnewline
100 & \$75.0M & \$37.70 & \$112.70\tabularnewline
300 & \$225M & \$5.36 & \$230.36\tabularnewline
600 & \$450M & \$0.29 & \$450.29\tabularnewline
1,000 & \$750M & \$0.01 & \$750.01\tabularnewline
3,000 & \$2,250M & \$0.00 & \$2,250.00\tabularnewline
6,000 & \$4,500M & \$0.00 & \$4,500.00\tabularnewline
10,000 & \$7,500M & \$0.00 & \$7,500.00\tabularnewline
\hline 
\end{tabular}
\par\end{centering}

\caption{Prize fund total ($r_{l}NPt$), future value per share ($V_{future}$),
and expected value per share ($V_{expected}$) for the SETI Lottery
Bond, assuming $N=10^{6}$ as the number of shares sold, $P=\$100$
as the price per share, $r_{c}=2\%$ as the bond coupon rate, and
$r_{l}=0.75\%$ as the contribution toward the prize fund. Future
values are adjusted for inflation according to Eq. (\ref{eq:FV}),
assuming an inflation rate of $r_{i}=3\%$. The prize fund is assumed
to grow at a constant rate equal to the rate of inflation. Under these
assumptions, the future value falls with time because its growth is
less than inflation, but the expected value of each share rises with
time as the prize fund grows according to Eq. (\ref{eq:EV}). Note
that the expected value of a share exceeds its purchase price after
about sixty years. \label{tab:Prices}}
\end{table}
, which assumes $N=10^{6}$ as the number of shares sold, $P=\$100$
as the price per share, $r_{c}=2\%$ as the bond coupon rate, $r_{l}=0.75\%$
as the contribution toward the prize fund, and $r_{i}=3\%$ as the
rate of inflation. Calculations in Table \ref{tab:Prices} assume
that the lottery prize fund is managed so as to keep pace with inflation.
Although the future value of each share decreases with time, the expected
value exceeds the purchase price after about sixty years. During the
first ten years of the SLB, an investor will see a modest decrease
in future value but may remain hopeful of winning a modest prize if
SETI succeeds. After about fifty to one hundred years, an investor's
share will decrease notably in future value due to inflation, but
the expected value of the share is now much greater, and SLB shares
may therefore be valuable heirlooms that are passed to kin or benefacted
to charity. On even longer timescales that span generations, SLB shares
would have very low future value, due to inflation, but would have
a very large expected value if SETI succeeds and the lottery commences.
If SETI takes hundreds or even thousands of years to succeed, then
SLB shares may become treasures that are cherished and guarded by
families and charities, both as a relic from the past and a hope for
a prosperous financial future.

In order to remain viable, the growth of the SLB fund must be sufficient
to satisfy obligations to investors, the SETI community, and the lottery
prize fund. Additionally, the total value of the fund must always
be at least equal to the principal investment $NP$, while the lottery
fund and the annual SETI contribution must grow to keep pace with
inflation. Assuming annual compounding, this can be expressed as

\begin{equation}
NP\left(\frac{1+R}{1+r_{i}}\right)^{t}\geq NP+r_{c}NPt+\left(r_{l}+r_{s}\right)\left(1+r_{i}\right)^{t}NPt,\label{eq:inequality}
\end{equation}
where $R$ is the required growth rate of the SLB fund. Solving for
$R$ yields the equation 

\begin{equation}
R\ge\exp\left\{ \frac{1}{t}\ln\left[1+t\left(r_{c}+\left(r_{l}+r_{s}\right)\left(1+r_{i}\right)^{t}\right)\right]\right\} \left(1+r_{i}\right)-1,\label{eq:R}
\end{equation}
that must be evaluated numerically. Table \ref{tab:GrowthRates}
\begin{table}
\begin{centering}
\begin{tabular}{cccc}
\hline 
\multirow{3}{*}{\textbf{Time (yr)}} & \multicolumn{3}{c}{\textbf{Required Growth Rate for the SETI Lottery Bond Fund}}\tabularnewline
 & Conservative & Moderate & Aggressive\tabularnewline
 & \$2.5M/yr for SETI & \$10M/yr for SETI & \$30M/yr for SETI\tabularnewline
\hline 
1 & 8.51\% & 16.5\% & 37.7\%\tabularnewline
3 & 8.43\% & 15.6\% & 31.2\%\tabularnewline
6 & 8.32\% & 14.5\% & 25.8\%\tabularnewline
10 & 8.20\% & 13.5\% & 21.8\%\tabularnewline
30 & 7.84\% & 11.0\% & 14.5\%\tabularnewline
60 & 7.59\% & 9.54\% & 11.4\%\tabularnewline
100 & 7.40\% & 8.66\% & 9.79\%\tabularnewline
300 & 6.90\% & 7.33\% & 7.70\%\tabularnewline
600 & 6.62\% & 6.83\% & 7.01\%\tabularnewline
1,000 & 6.46\% & 6.59\% & 6.70\%\tabularnewline
3,000 & 6.25\% & 6.29\% & 6.33\%\tabularnewline
6,000 & 6.18\% & 6.20\% & 6.22\%\tabularnewline
10,000 & 6.15\% & 6.16\% & 6.17\%\tabularnewline
\hline 
\end{tabular}
\par\end{centering}

\caption{Required growth rates ($R)$ and annual contribution to SETI research
($r_{s}NP$) for the SETI Lottery Bond, assuming $N=10^{6}$ as the
number of shares sold, $P=\$100$ as the price per share, $r_{c}=2\%$
as the bond coupon rate, and $r_{l}=0.75\%$ as the contribution toward
the prize fund. Growth rates are adjusted for inflation according
to Eq. (\ref{eq:R}), assuming an inflation rate of $r_{i}=3\%$.
The annual contribution toward SETI is also adjusted for inflation
according to Eq. (\ref{eq:inequality}). Any of these three growth
scenarios would provide a stream of income, similar to an endowment,
to support SETI research. \label{tab:GrowthRates}}
\end{table}
 shows values for this fund growth rate $R$ under three scenarios
labeled ``conservative'', ``moderate'', and ``aggressive'',
which illustrate configurations of the SLB that respectively yield
\$2.5 million, \$10 million, or \$30 million per year for SETI research.
Upon reaching maturation, the SLB lottery would commence by randomly
selecting a fraction of shares that each win a portion of the prize
fund. The fraction of shares that win a prize should be small enough
that individual prizes are desirable but large enough that investors
are attracted by the chances of winning. Even if one out of every
ten thousand shares were selected for a prize, this would produce
several winners who each make a handsome profit compared to their
initial investment. The SLB thus serves as a long-term savings product
with a continually growing lottery prize in anticipation of the first
discovery of ETI.

The values in Tables \ref{tab:Prices} and \ref{tab:GrowthRates}
are calculated out to ten thousand years, but no currency has a legacy
anywhere near this length. Within a few hundred years, the U.S. dollar
may be surpassed by the euro or yuan as the world ``reserve currency''
{[}21, 22{]}, while in a few thousand years, none of these units of
currency may be in use. Concepts of economics and value may even change
so drastically over thousand-year time scales that savings products
such as bonds or lotteries are no longer offered or wildly different
in form. Without the ability to predict how future economics will
develop on thousand-year timescales, the very long-term prices and
rates shown in Tables \ref{tab:Prices} and \ref{tab:GrowthRates}
should be considered illustrative at best.

Creation of the SLB will require cooperation with a financial institution
that has the legal authority to engage in banking and gaming activities.
Banking and gaming are regulated industries in most nations, with
about two thirds of lotteries operated by private or government-owned
corporations and the remaining third directly by government agencies
{[}9{]}. Financial institutions such as banks or large investment
corporations can also manage a large pool of funds to yield a profit,
for example by creating loans or engaging in other risky investments.
In order to succeed, the SLB fund must exhibit steady growth in order
to satisfy its obligations; financial institutions can generate this
steady growth of the principal investment and also can absorb any
losses that may occur during the fund's inception. Note that each
of the three scenarios in Table \ref{tab:GrowthRates} requires a
relatively high growth rate during the first ten to one hundred years
but a much lower rate after. While it might be difficult for an individual
investor to achieve this sort of performance, a large and viable financial
institution reasonably could manage the SLB fund to reach these goals.
A financial institution would be motivated to issue such a product
because any performance of the fund above expectations would yield
profit for the fund's managing institution (although any below-expected
performance would result in a loss for the financial institution),
which would provide a novel form of revenue from a new group of investors.
I therefore propose that the SLB should be a savings product issued
by a financial institution (such as a bank or a government treasury
department) and managed for the benefit of investors and the SETI
community.

SLB shares are also transferable, meaning that they can change ownership
and be passed from one generation to another. This lends to the creation
of a secondary market for trading SLB shares, and it seems likely
that many investors will buy or sell their SLB shares in such a market
well in advance of the discovery of ETI. In fact, the market value
of SLB shares may even ebb and flow with current events, such as the
discovery of new habitable planets or budget shortfalls that leave
important missions unfunded. This secondary market may help to maintain
interest in the SLB by providing new means for clever traders to reap
a profit. The transferability of SLB shares also make them attractive
intergenerational investments for families or organizations that want
to provide a fixed stream of income for the future. 

The SLB will generate a regular stream of income for the SETI community
that can be distributed to institutions and individuals to pursue
the search for extraterrestrial intelligent life. The specific means
of distribution of these funds likely will be a contentious issue
but ultimately will remain the decision of the financial institution
providing the SLB product. One possibility is direct endowment to
an existing SETI research facility, such as the SETI Institute in
Mountain View, California. The SETI Institute has been struggling
to support the Allen Telescope Array (ATA)---a radio interferometer
designed specifically for SETI---which requires about \$2.5 million
per year for its operations and science campaign, while a larger budget
would allow planned expansion of the ATA to continue. Any of the three
scenarios in Table \ref{tab:GrowthRates} provide sufficient capital
for SETI to revitalize the ATA, so direct endowment to existing SETI
organizations may be prudent. The sponsoring financial institution
could also solicit SETI researchers for grant proposals as a way of
distributing some of the funds, which may encourage new SETI research
groups to form in universities across the world. Banks and other financial
institutions may lack the expertise to conduct scientific review panels,
so partnerships with scientific organizations and foundations may
be prudent for the evaluation of proposals and dispersal of funds.
By supporting existing research efforts, revitalizing existing technology,
and providing new funding opportunities for investigators, the SLB
could provide tremendous opportunity to further SETI research.

Shares of the SLB can be purchased any time until the first discovery
of ETI, culminating in the lottery and repayment of the principal
investment. Because the first discovery of ETI will likely be contentious
and require a series of attempts at confirmation, I propose that an
independent scientific governance board should be tasked with determining
when the discovery and confirmation of extraterrestrial intelligent
life has occurred. The scientific governance board has the authority
to suspend the sale of SLB shares, including a freeze on trades in
any secondary markets, in order to evaluate a potential SETI discovery
for its authenticity. Once potential claims have been evaluated, the
scientific governance board can then decide to re-open the market
(due to a false positive) or close the market and begin the lottery
(due to the discovery of ETI). The scientific governance board should
include global representation of scientists in the SETI and astrobiology
communities as well as scholars from other academic disciplines, all
of whom are financially independent of the SLB. Some representation
from other experts also may be prudent; however, the purpose of the
governance board is solely to evaluate whether or not enough information
exists to say that the discovery of extraterrestrial intelligent life
has occurred. Presumably, any reported discovery of ETI will be thoroughly
vetted by the scientific community, and it is not the job of the scientific
governance board to conduct follow-up observations. Instead, the scientific
governance board serves as a mechanism to regulate the closing of
the SLB market when the discovery of ETI eventually occurs.

In summary, the SETI Lottery Bond is a savings product issued by a
financial institution that provides a fixed rate of return to the
investor, a fixed rate of return to support SETI research, and a chance
at winning a prize through lottery when the bond matures due to the
discovery of ETI.

\section{Ethical Considerations\label{sec:Ethical-Considerations}}

The SETI Lottery Bond is a savings product issued by a financial institution
that provides continuous funding for SETI research. This product involves
a conflation of investment and scientific goals, which raises several
ethical issues that are worth addressing in advance.

First, if the SLB has a duration that lasts until the first discovery
of ETI, then what, exactly, constitutes a ``discovery of extraterrestrial
intelligent life''? Direct contact between humans and ETI seems unlikely---and
SETI efforts are unlikely to alter this probability, anyhow. Any evidence
of ETI will therefore be subject to some analysis and interpretation,
which may cause a lengthy period of discussion within the scientific
community, probably followed by an even longer period of debate within
public, political, and religious circles. For the purposes of the
SLB, evidence of ETI could take the form of narrow-band radio signals
emanating from an ETI civilization that show patterns or encoded information
{[}1{]}, functional or defunct technological artifacts sent to the
Solar System by ETI that may be drifting nearby {[}2, 3{]}, or large
excesses of infrared radiation from stars, clusters, or galaxies that
indicate macro-engineering feats of an ETI empire {[}4{]}. Any of
these discoveries would provide extraordinary evidence for ETI and
would become targets for further exploration or attempts at communication.
Authenticating the signal or artifact will be an arduous process that
may take years or decades before scientific consensus. For this reason,
the SLB stipulates a scientific governance board with the authority
to suspend transactions if the authenticity of an ETI discovery is
under scrutiny. Any discovery of ETI must broadly be considered as
scientific consensus in order to constitute the first ``discovery
of extraterrestrial intelligent life'' that would bring the SLB to
maturity.

Many nations have explicit laws prohibiting insider trading of public
securities, and the SLB should be no different. In particular, the
SLB provides financial support for SETI research, which will directly
benefit a small number of scientists worldwide. This creates the potential
for insider trading activities because SETI scientists may have access
to privileged information, which may provide an unfair advantage when
trading SLB shares in a secondary market. Therefore, a provision against
insider trading should also include a restriction that anyone who
receives funding directly from the SLB is prohibited from personally
investing in the fund.

Another consideration is that a long-term fund like the SLB might
be assumed to be a sort of Ponzi scheme or other fraudulent investment.
In a Ponzi scheme, returns to investors are payed either from preexisting
coffers or from the contributions of future investors, which requires
a continuous supply of revenue and almost always fails in collapse.
Ponzi schemes are not true savings products because they falsely advertise
a rate of return that cannot be honored for all investors. The SLB,
contrary to any fraudulent investment scheme, is a legitimate financial
savings product, issued by an authorized financial institution, that
promises a fixed rate of return to investors as well as the return
of principal upon reaching maturity. Because maturity occurs upon
the discovery of ETI, likely to be a long time from the bond's inception,
the SLB fund managers must be careful to maintain the principal investment
balance for when maturity occurs---otherwise, a shortage of funds
upon the discovery of ETI would result in a failure to meet investor
obligations. Managed properly by an authorized financial institution,
the SLB would act as a novel savings product with a very long expected
duration.

The inclusion of a lottery prize in the SLB raises the issue of whether
or not a savings product should include random elements. In particular,
lotteries and other games of chance seem to reinforce gambling behavior
patterns, even when used to fund programs for community development
{[}16, 17{]}. However, products such as lottery-linked deposit accounts
or prize-linked savings accounts appear to successfully attract individuals
otherwise likely to gamble by providing a product that allows savings
to build but with a randomized rate of return that captures the ``thrill''
of a lottery {[}9, 10, 12, 13{]}. Lottery bonds and lottery-linked/prize-linked
accounts are primarily a marketing device to appeal to investors who
enjoy purchasing products with unknown rates of return, and they are
potentially more attractive than conventional gaming products because
they promise a full return of the principal investment. Although a
lottery bond does include elements of gaming, such a product incorporates
a ``clear savings element'' {[}11{]} with ``entertainment value''
{[}9{]} that distinguishes itself from conventional gaming products.

It is impossible to predict the demographics of SLB investors, but
I suspect that the SLB would appeal to a wide range of socioeconomic
classes, perhaps wider than most existing lottery or lottery-linked
products. With the SLB, the ``thrill'' of investment is the knowledge
that an investor is helping to further scientific progress in SETI
and the anticipation of a lottery prize if the search succeeds. This
type of marketing may appeal to science enthusiasts, altruists, futurists,
and ``ethical investors'' who seek to help SETI but are uninterested
in purely charitable donations {[}9{]}. Furthermore, if SLB shares
are offered at a relatively affordable price (such as \$100 per share
as assumed in Table \ref{tab:Prices}), then individuals of nearly
any economic class will have the option of participating---and if
the SLB shows any similarity to existing lottery-linked products {[}12,
13, 20{]}, then it may be help to discourage gambling and encourage
long-term savings. Analysis by {[}20{]} suggests that more than half
of individuals in the U.S. would consider purchasing a lottery-linked
product, with primary interest among individuals without savings habits.
The SLB may represent an opportunity for financial institutions to
draw consumers from an otherwise untapped market.

Lottery-linked savings products could be useful in funding a variety
of research or causes, and it is reasonable to question why this proposal
calls for SETI lottery bonds in particular, rather than any other
charitable cause. While a device like the SLB certainly could be used
to finance a variety of research projects (such as nuclear fusion
development, or the construction of large particle accelerators, or
applied medical research), three aspects of SETI in particular make
it a prime candidate for associating with a lottery-linked product.
First, SETI represents a search with no known duration: it could take
ten years, a hundred years, or a thousand or more before SETI finds
signs of ETI, and calculating the expected amount of time before SETI's
success is a challenging task to say the least. Compared to most other
areas of research, SETI is one of the few where the amount of time
required for success is probably on the order of generations. The
SLB is constructed so that the date of maturity is likewise unknown,
allowing investors to participate in the uncertainty that accompanies
SETI. Second, SETI is a long-term effort that requires a continuous
stream of funding if it is ever to succeed. Appealing to individual
donors time after time, each of whom may lose interest after a few
decades of null results, has proven to be an ineffective strategy
in sustaining SETI, so a perpetual bond is an option for providing
a sustained source of revenue. Third, SETI efforts are admittedly
optimistic and sometimes described as a gamble with unknown odds;
in a way, the thrill of winning a lottery prize in the SLB allows
investors to share in SETI scientists' hopes of eventually discovering
ETI---if and when this momentous occasion occurs, scientists and the
world will celebrate, but SLB investors will also partake finally
in their long overdue lottery. Other communities certainly could implement
lottery linked products tailored to their research; the SLB represents
an example of such a product designed for SETI in particular.

A final objection, perhaps mostly from the scientific community, is
that this fundamentally is not the way science should be financed.
For most of history, scientists have relied upon the goodwill of wealthy
patrons and governments to pursue basic research, which has caused
funding to ebb and flow with the tides of politics. Today, the vast
majority of science funding takes the form of competitive grants offered
by various government organizations, while private organizations offering
support for science remain a much smaller contributor. Some research
by for-profit corporations also leads to new basic scientific knowledge,
but these results are not always published in peer-reviewed journals
and are sometimes even guarded as trade secrets. Many scientists in
academia tend to frown upon linking basic research with any profit
motive, believing that science should be guided only by motives such
as curiosity or passion. Indeed, the profit motive introduced by the
SLB might increase the likelihood of a false positive discovery of
ETI---perhaps because any investors in the SLB will have a personal
attachment to the outcome of SETI. This highlights the need for a
scientific governance board to accompany any implementation of the
SLB and be particularly wary of false positives that might arise early
on. Nevertheless, ideological objections to the SLB remain a valid
point of contention, and many researchers will likely avoid any funding
linked to a financial product. Even so, some scientists---either out
of necessity, invention, or out of frustration with the current grant-based
system---may find the SLB to be an appealing source of funding, one
that allows investors (instead of just donors) to engage in science.
I do not suggest that this should be the dominant model for funding
in science, nor do I argue that this model is better (or worse) than
current models of funding. My goal in crafting this proposal is to
suggest a way that SETI in particular could harness the capital it
needs to sustain the long-term search for our extraterrestrial neighbors.
If grants and donations cannot achieve this goal, then why not try
a SETI Lottery Bond?

\section{Conclusion}

I have outlined a proposal for the SETI Lottery Bond as a fixed rate
perpetual bond with a lottery at maturation that can serve as a continuous
source of funding to sustain the search for extraterrestrial intelligent
life. The SLB is a savings product issued by a financial institution
and would appeal to people interested in assisting SETI though investment,
rather than donation. Because SETI is a long-term endeavor, one that
will likely continue to span generations, the SLB is a long-term savings
product that will likewise span generations, which may encourage intergenerational
wealth building among individuals without savings habits. Regular
payments from the SLB will sustain SETI research until the first discovery
of ETI by providing several to tens of million dollars in funding
annually. Funding at this level will allow SETI surveys, such as the
Allan Telescope Array, to remain dedicated to SETI and expand their
capabilities as necessary, which is a necessary first step to find
evidence of ETI.

Bringing the SLB into reality presents a challenge of convincing a
financial institution to seek authorization to offer such a lottery
bond as a product. Many European nations and developing nations offer
lottery bonds or lottery-linked savings accounts {[}10, 11, 12, 13{]},
and evidence suggests that consumers in the U.S. are interested in
lottery-linked products {[}20{]}. A bank in the U.S., for example,
would be required to secure federal and state approval before it could
begin issuing lottery bonds, such as the SLB, as such a product includes
elements of both banking and gaming. While securing this approval
might be a tedious process, the SLB might ultimately be approved as
a savings product because the primary goal of the SLB is to capture
investor zeal to raise capital for SETI research. Savings products
offered by banks are usually not linked to altruistic objectives,
and almost never linked to unknown-risk outcomes such as SETI, so
the SLB represents a unique product that could be offered by any stable
financial institute willing to offer a long-term perpetual bond that
lasts until SETI succeeds. Perhaps offering such a product would also
help financial institutions to improve their perception among consumers.

SETI is a global endeavor, and the SLB may be most successful at a
global scale. Analysis by {[}9{]} discusses the possibility of implementing
a global lottery bond and claims that such a product is ``best provided
by a single organization selling and administering the bonds worldwide''.
A single global bank, group of international banks, multinational
investment corporation, or consortium of international governments
would therefore be among the best options for offering the SLB worldwide.
Products like the SLB could widen the range of savings products available
in developing nations {[}9{]}, and organizations like the World Bank
or the International Monetary Fund (IMF) might even partner with the
SLB managers in order to make the product available to nations with
few available savings options. 

Given that federal budgets are uncertain and already saturated, I
propose that one or more international financial institutions obtain
authorization to offer lottery-linked savings products in support
of SETI research. Even if SETI takes a very long time to succeed---or
if it never succeeds---this plan still will help to promote personal
savings and intergenerational wealth building along with astrobiology
education. In the best case, SETI succeeds; but even in the worst
case, if SETI finds nothing, then it will at least have taught us
how to think more seriously about our financial future.

\subsection*{Acknowledgments}

I thank Jill Tarter, Robin Hanson, Ravi Kopparapu, Seth Baum, Sanjoy
Som, Michael Busch, Brendan Mullan, and Biff Robillard for helpful
suggestions and comments. Any opinions are those of the author alone.
This research did not receive any specific grant from funding agencies
in the public, commercial, or not-for-profit sectors.

\subsection*{References}

\noindent {[}1{]} G. Cocconi, P. Morrison, Searching for interstellar
communications, Nature 184 (1959) 844\textendash 846.\\

\noindent {[}2{]} R.N. Bracewell, Communications from superior galactic
communities, Nature 186 (1960) 670\textendash 671.\\

\noindent {[}3{]} C. Rose, G. Wright, Inscribed matter as an energy-efficient
means of communication with an extraterrestrial civilization, Nature
431 (2004) 47\textendash 49.\\

\noindent {[}4{]} F.J. Dyson, Search for artificial stellar sources
of infrared radiation, Science 131 (1960) 1667\textendash 1668.\\

\noindent {[}5{]} J. Tarter, The search for extraterrestrial intelligence
(SETI), Annual Review of Astronomy \& Astrophysics 39 (2001) 511\textendash 548.\\

\noindent {[}6{]} I. Almár, J. Tarter, The discovery of ETI as a high-consequence,
low-probability event, Acta Astronautica 68 (2011) 358\textendash 361.\\

\noindent {[}7{]} M. Dominik, J.C. Zarnecki, The detection of extra-terrestrial
life and the consequences for science and society, Philosophical Transactions
of the Royal Society A: Mathematical, Physical and Engineering Sciences
369 (2011) 499\textendash 507.\\

\noindent {[}8{]} T. Peters, The implications of the discovery of
extra-terrestrial life for religion, Philosophical Transactions of
the Royal Society A: Mathematical, Physical and Engineering Sciences
369 (2011) 644\textendash 655.\\

\noindent {[}9{]} T. Addison, A.R. Chowdhury, A global lottery and
a global premium bond, WIDER Discussion Papers, World Institute for
Development Economics (UNU- WIDER), No. 2003/80, 2003.\\

\noindent {[}10{]} S. Lobe, A. Hölzl, Why are British Premium Bonds
so Successful? The Effect of Saving With a Thrill, doi:10.2139/ssrn.992794,
2007.\\

\noindent {[}11{]} P. Tufano, Saving whilst gambling: An empirical
analysis of UK premium bonds, American Economic Review: Papers \&
Proceedings 98 (2008) 321\textendash 326.\\

\noindent {[}12{]} M.F. Guillén, A.E. Tschoegl, Banking on gambling:
Banks and lottery-linked deposit accounts, Journal of Financial Services
Research 21 (2002) 219\textendash 231.\\

\noindent {[}13{]} M.S. Kearney, P. Tufano, J. Guryan, E. Hurst, Making
savers winners: An overview of prize-linked savings products, in:
O.S. Mitchell and A. Lusardi (Eds.), Financial Literacy: Implications
for Retirement Security and the Financial Marketplace, 2011, Oxford
University Press, pp. 218\textendash 240.\\

\noindent {[}14{]} J. Harris, The survival lottery, Philosophy 50
(1975) 81\textendash 87.\\

\noindent {[}15{]} G.J. Annas, The prostitute, the playboy, and the
poet: Rationing schemes for organ transplantation, American Journal
of Public Health 75 (1985) 187\textendash 189.\\

\noindent {[}16{]} M.O. Borg, P.M. Mason, The budgetary incidence
of a lottery to support education, National Tax Journal 41 (1988)
75\textendash 85.\\

\noindent {[}17{]} R. Rubenstein, B. Scafidi, Who pays and who benefits:
Examining the distributional consequences of the Georgia lottery for
education, National Tax Journal 55 (2002) 223\textendash 238.\\

\noindent {[}18{]} H. Lévy-Ullmann, Lottery bonds in France and in
the principal countries of Europe, Harvard Law Review 9 (1896) 386\textendash 405.\\

\noindent {[}19{]} R.C. Green, K. Rydqvist, Ex-day behavior with dividend
preference and limitations to short-term arbitrage: the case of Swedish
lottery bonds, Journal of Financial Economics 53 (1999) 145\textendash 187.\\

\noindent {[}20{]} P. Tufano, J.-E. De Neve, N. Maynard, U.S. consumer
demand for prize-linked savings: New evidence on a new product, Economics
Letters 111 (2011) 116\textendash 118.\\

\noindent {[}21{]} B. Eichengreen, Sterling's past, dollar's future:
Historical perspectives on reserve currency competition, NBER Working
Papers, No. w11336, 2005.\\

\noindent {[}22{]} M. Chinn, J.A. Frankel, Will the euro eventually
surpass the dollar as leading international reserve currency?, G7
Current Account Imbalances: Sustainability and Adjustment, 2007, National
Bureau of Economic Research, pp. 283\textendash 338.
\end{document}